\documentclass{article}                     

\usepackage{amsmath,amssymb,graphicx,color,relsize,here}

\newtheorem{corollary}{\bf Corollary}[section]

\usepackage{tikz}
\usetikzlibrary{arrows,calc}
\usetikzlibrary{shapes,snakes}
\usetikzlibrary{mindmap,trees,backgrounds}
\usepackage{bm}
\usetikzlibrary{scopes}
\usepackage{lineno,hyperref,amsmath,amssymb,pgfplots}
\usepackage{pgf,tikz}
\usetikzlibrary{arrows}

\newtheorem{assumption}{\bf Assumption}
\newtheorem{proposition}{\bf Proposition}

\usepackage{graphicx}
\begin{document}

\title{NLP Solutions as Asymptotic Values of ODE Trajectories}
\author{Mazen Alamir \ \\ \ \\  CNRS/Gipsa-lab Control systems department.\\ Email : mazen.alamir@grenoble-inp.fr}


\date{}

\maketitle

\begin{abstract}
\noindent In this paper, it is shown that the solutions of general differentiable constrained optimization problems can be viewed as asymptotic solutions to  sets of Ordinary Differential Equations (ODEs). The construction of the ODE associated to the optimization problem is based on an exact penalty formulation in which the weighting parameter dynamics is coordinated with that of the decision variable so that there is no need to solve a sequence of optimization problems, instead, a single ODE has to be solved using available efficient methods. Examples are given in order to illustrate the results. This includes a novel systematic approach to solve combinatoric optimization problems as well as fast computation of a class of optimization problems using analogic circuits leading to fast, parallel and highly scalable solutions.
\end{abstract}

\section{Introduction}
\label{secintro}
Consider the following optimization problem with inequality constraints:
\begin{eqnarray}
\min_{x\in \mathbb{R}^{n}} f(x) \ \mbox{\rm under}\ c_i(x)\le 0\quad i\in \{1,\dots,n_c\} \label{defdeP} 
\end{eqnarray} 
where $f$ and $c_i$'s are scalar functions of the decision variable $x\in \mathbb{R}^{n}$. Let $f$ be the exact penalty induced function defined by:
\begin{eqnarray}
\bar f(x,\rho)=f(x)+\rho\cdot \psi(x) \label{defdefxrho} 
\end{eqnarray} 
where $\psi(x)$ is given by:
\begin{eqnarray}
\psi(x):=\sum_{i=1}^{n_c}\bigl[\max\{0,c_i(x)\}\bigr]^m\quad ;\quad m\in \mathbb{N} \label{defdepsi} 
\end{eqnarray} 
A wide class of algorithms intends to solve (\ref{defdeP}) by solving a sequence of unconstrained optimization problems of the form
\begin{eqnarray}
\min_{x\in \mathbb{R}^n} \bar f(x,\rho_k)
\end{eqnarray} 
for a varying (generally increasing) values of the weighting coefficient $\rho_k$. For each problem in the sequence, only $x$ is searched for while $\rho_k$ is kept constant \cite{Byrd:99,Birgin:2012}. The series of unconstrained problems are sometimes replaced a series of problems with by box constraints as descent method with projection are easy to perform. \ \\ \ \\ 
The increase of $\rho$ is generally defined by $\rho_{k+1}\leftarrow r\times \rho_k$ with $r>1$. The sequence of precision parameter $\omega_k$ (to which the intermediate problems have to be solved) is also made such that $\omega_k\rightarrow 0$ in order to avoid solving with a uselessly high precision intermediate problems. It is then obvious that the efficiency of the resulting algorithms is tightly related to the choice of $r$ and the intermediate precision sequence $\{\omega_k\}_k$ since small values of $r>1$ leads to unnecessarily high number of intermediate problems while a too high values of $r>1$ leads to stiff problems that may lead to slow convergence (because this makes the solution of the box-constraint subproblem harder \cite{Birgin:2012}) beside the fact that it breaks the {\em continuation} argument that underlies the whole scheme. The choice of $\omega_k$ corresponds to similar trade-offs that need to be carefully handled. Such issues are extensively studied in \cite{Birgin:2012} leading to non necessarily monotonic behavior of $\rho_k$ when solving the sequence of intermediate box constrained modified Lagrangian problems in order to avoid high number of iterations that result when $\rho_k$ is unnecessarily high. This recent study \cite{Birgin:2012} shows at least that monitoring the dynamic evolution of $\rho_k$ is not a trivial issue.\ \\ \ \\ 
In this paper, it is shown that simultaneous dynamics of $x$ and $\rho$ can be defined through a differential equation of the form:
\begin{eqnarray}
\dot x=F_1(x,\rho)\ ;\ \dot\rho=F_2(x,\rho) \label{eqdiff} 
\end{eqnarray} 
such that solving (\ref{eqdiff}) gives trajectories that asymptotically converge towards the set of solutions of (\ref{defdeP}). \ \\ \ \\ 
Note that by doing so, the present paper does not propose a specific alternative algorithm to solve the NLP problem (\ref{defdeP}). Rather, it enables all the efficient algorithms that are available through the huge literature on ODE integration to become candidate algorithms for (\ref{defdeP}). Moreover all the computational background regarding many issues (such as parametric sensitivity \cite{Leis:1988}, parallel computing \cite{Voss199765},  precision monitoring to cite but few items) become  available for the constrained optimization problem paradigm. As such, the result of the present paper can be viewed as a starting point for future investigation. More interestingly, it is shown briefly in this paper that expressing the fact that solving (\ref{defdeP}) can be done by integrating ODEs enables (for some specific problems) to built analogic circuits that can achieve the task in fast, parallel and massively scalable way. \ \\ \ \\ 
This paper is organized as follows: first section \ref{secdefnot} gives the definitions and notation used throughout the paper. Section \ref{secworkingassumptions} states the working assumptions that are needed to derive the main results of the paper. These results are stated and proved in section \ref{secmainresult} while section \ref{secexamples} gives some examples of application of the paper results. Finally the paper ends with section \ref{secconclusion} that summarizes the contribution and gives hints for further investigations.    
\subsection{Definition \& Notation} \label{secdefnot} 
\noindent Throughout the paper, the following notation is used. The $n$-dimensional vectors $f_x(x)$, $\psi_x(x)$ and $\bar f_x(x,\rho)$ denote the gradients of $f$, $\psi$ and $\bar f$ w.r.t $x$. The scalar function $g(x,\rho)$ denotes the euclidian norm of $\bar f_x$ at $(x,\rho)$, namely:
\begin{eqnarray}
g(x,\rho):=\|\bar f_x(x,\rho)\|
\end{eqnarray} 
For a given weighting coefficient $\rho>0$, the set of stationary points of $\bar f(\cdot,\rho)$ is denoted by $\mathcal S_\rho$, namely:
\begin{eqnarray}
\mathcal S_\rho:=\bigl\{x\in \mathbb{R}^{n}\ \vert\  g(x,\rho)=0\bigr\}
\end{eqnarray} 
For any $x\in \mathbb{R}^{n}$, the notation $d(x,\rho)$ refer to the distance between $x$ and the set $\mathcal S_\rho$, namely:
\begin{eqnarray}
d(x,\rho):=\min_{z\in \mathcal S_\rho}\|x-z\|
\end{eqnarray} 
The set of admissible values of $x$ is denoted by:
\begin{eqnarray}
\mathcal A:=\bigl\{x\in \mathbb{R}^{n}\ \vert\ \psi(x)=0\bigr\}
\end{eqnarray} 
\section{Working Assumptions} \label{secworkingassumptions} 
\noindent The first assumption states that the optimization problem is well posed in the sense that either the original cost $f(x)$ is already lower bounded or the constraints are such that the weighted cost $\bar f$ is lower bounded:\\
\begin{assumption}{\bf [Well posedness]}\label{ass1} 
For any $\rho>0$, there is a lower bound $\bar f_{\min}(\rho)$ such that $\bar f(x,\rho)\ge \bar f_{min}(\rho)$ for all $x\in \mathbb{R}^{n}$.
\end{assumption}
\ \\
Note that in the framework of the present paper, it is not assumed that the functions involved are convex. This means that the set $\mathcal S_\rho$ may not be a singleton $\{x^*(\rho)\}$, it is assumed that the norm of the gradient of the weighted cost $\bar  f(\cdot,\rho)$ {\bf away} from $\mathcal S_\rho$ can be bounded below by the distance to the set $\mathcal S_\rho$ through some coefficient $k_c$. This leads to the following generalization of the strong convexity assumption:\\
\begin{assumption}{\bf [$\mathcal (S_\rho$)-Strong Convexity]} \label{ass2} 
There is a constant $k_c>0$ such that the following inequality holds:
\begin{eqnarray}
g(x,\rho)\ge k_c\times d(x,\rho) \label{defdekc} 
\end{eqnarray} 
for all $x\in \mathbb{R}^{n}$.
\end{assumption}
\ \\
Note that contrary to the classical strong convexity assumption that involves two arbitrary points $x_1$ and $x_2$, the inequality (\ref{defdekc}) involves the distance from an arbitrary $x$ to those points lying inside the set of stationary points $\mathcal S_\rho$.\ \\ \ \\ 
The Next assumption describes a generalized Lypschitz-like assumption on the constraints and the way they are used to construct the exact penalty term $\psi(x)$. \\
\begin{assumption}{\bf [Growth rate of $\psi$]} \label{ass3}  There is a polynomial $P$ of degree $n_\psi\in \mathbb{N}$ with $P(0)=0$  that satisfies the  following inequality
\begin{eqnarray}
\vert\psi(x_2)-\psi(x_1)\vert \le P(\|x_2-x_1\|)\bigr]\label{defdegrowthpsi} 
\end{eqnarray} 
for all $x\in \mathbb{R}^{n}$. 
\end{assumption}
\ \\
Note that the use of the polynomial $P$ of the form:
\begin{eqnarray}
P(d):=\sum_{i=1}^{n_\psi} \alpha_i d^i \label{defdePol} 
\end{eqnarray} 
accounts for the possibility to use different penalty exponents $m$ in the definition of the constraint penalty term in (\ref{defdepsi}) and the fact that the bounding function may involve lower powers for small distances $d$ and higher powers far from the set $\mathcal A$. \ \\ \ \\ 
The following assumption is needed to guarantee the existence of solutions to the ODE built up with the functions $\bar f_x$ and $\psi$:\\
\begin{assumption}{\bf [Locally-Lypschitz maps]} \label{ass4} 
For all finite $\rho>0$ the maps $f_x(\cdot,\rho)$, $\psi(\cdot)$ and $\psi_x(\cdot)$ are {\bf locally} Lypschitz.
\end{assumption}
\ \\
Note that this last assumption expresses {\em local} requirement while (\ref{defdekc}) and (\ref{defdegrowthpsi}) are required to hold for any $x$.\ \\ \ \\ 
The last assumption concerns the relevance of the use of the penalty method to solve (\ref{defdeP}). It states that when the penalty coefficient $\rho$ goes to infinity, the  possible stationary points for the weighted cost converge towards the admissible set $\mathcal A$:\\
\begin{assumption}{\bf [Relevance of the penalty approach]}\label{ass5} 
\begin{eqnarray}
\lim_{\rho\rightarrow \infty}\Bigl[\ \sup_{x\in \mathcal S_\rho}\psi(x)\ \Bigr]=0 \label{eqass} 
\end{eqnarray} 
\end{assumption}
\ \\ 
This assumption is almost implicitly required in any penalty-based approach to solve the constrained optimization problem (\ref{defdeP}). It can obviously be replaced by some more apparently trivial assumptions that can be used to prove (\ref{eqass}). The short form is preferred here for the sake of clarity.  
\section{Main Results} \label{secmainresult}
The main result of the paper can be stated in the following proposition:\\
\begin{proposition}{\bf [Main Result]} \label{prop1} 
Assume that some $(\lambda,q)\in \mathbb{R}_+^*\times \mathbb{N}$ is chosen. Consider the following system of differential equations:
\begin{eqnarray}
\dot x&=&-\Bigl[\sum_{i=1}^{q}\dfrac{(\lambda\cdot g(x,\rho))^{i-1}}{(i-1)!}\Bigr]\times \bar f_x(x,\rho) \label{ode1}\\
\dot \rho &=& \gamma\times \psi(x) \label{ode2} 
\end{eqnarray} 
{\bf if} the following conditions hold:
\begin{enumerate}
\item Assumptions \ref{ass1}-\ref{ass5} are satisfied
\item $q\ge n_\psi$ [see (\ref{defdePol})]
\end{enumerate}
{\bf then} for any $\lambda>0$, there is a sufficiently small $\gamma>0$ such that any asymptotic solution of (\ref{ode1})-(\ref{ode2}) satisfies the KKT necessary conditions of optimality for the constrained optimization problem (\ref{defdeP}). $\hfill \heartsuit$
\end{proposition}
\ \\
{\sc Proof}. Let us compute the derivative of the weighted cost $\bar f(x,\rho)$:
\begin{eqnarray}
\dfrac{d\bar f}{dt}&=&\gamma\psi^2(x)-\Bigl[\sum_{i=1}^{q}\dfrac{(\lambda\cdot g(x,\rho))^{i-1}}{(i-1)!}\Bigr]^2\times g^2(x,\rho) \nonumber \\
&=&\gamma\psi^2(x)-\Bigl[\sum_{i=1}^{q}\dfrac{(\lambda\cdot g(x,\rho))^{i}}{\lambda\cdot (i-1)!}\Bigr]^2 \label{iju8} 
\end{eqnarray} 
Let $x_s(\rho)\in \mathcal S_\rho$ be the closest point to $x$ that lies inside the stationary set $\mathcal S_\rho$. According to (\ref{defdegrowthpsi}) of Assumption \ref{ass3}, one can write:
\begin{eqnarray}
\psi(x)&\le& \psi(x_s(\rho))+P(\|x-x_s(\rho)\|)\\
&\le& \psi(x_s(\rho))+P(d(x,\rho)) \label{tfr5} 
\end{eqnarray}  
Now by virtue of Assumption \ref{ass5}, $\psi(x_s(\rho))$ satisfies the following asymptotic property:
\begin{eqnarray}
\psi(x_s(\rho))=O(1/\rho)
\end{eqnarray} 
Therefore, (\ref{tfr5}) becomes [using (\ref{defdePol})] :
\begin{eqnarray}
\psi(x)&\le& P(d(x,\rho))+O(1/\rho)\\
&\le& \Bigl[\sum_{i=1}^{n_\psi}\alpha_id^i(x,\rho)\Bigr]+O(1/\rho) \label{pkhg58} 
\end{eqnarray} 
On the other hand, the sum in the r.h.s of (\ref{iju8}) satisfies [because of (\ref{defdekc})] the following inequality:
\begin{eqnarray}
\Bigl[\sum_{i=1}^{q}\dfrac{(\lambda\cdot g(x,\rho))^{i}}{\lambda\cdot(i-1)!}\Bigr]^2\ge \Bigl[\sum_{i=1}^{q}\dfrac{(\lambda\cdot k_c\cdot d(x,\rho))^{i}}{\lambda\cdot(i-1)!}\Bigr]^2 \label{tgFFFF} 
\end{eqnarray} 
Now using (\ref{pkhg58})  and (\ref{tgFFFF}) in (\ref{iju8}) enables to write [dropping all the terms with indices higher than $n_\psi\le q$ in the summing term of (\ref{iju8})] and using the identity $y_1^2-y_2^2=(y_1+y_2)(y_2-y_1)$:
\begin{eqnarray}
\dfrac{d\bar f}{dt}\le \Bigl[\sum_{i=1}^{n_\psi}\beta_i^+d^i+O(1/\rho)\Bigr]\cdot \Bigl[\sum_{i=1}^{n_\psi}\beta_i^-d^i+O(1/\rho)\Bigr] \label{POh6} 
\end{eqnarray} 
where $d^i:=d^i(x,\rho)$ while $\beta_i^+$ and $\beta_i^-$ are given by:
\begin{eqnarray}
\beta_i^+&=&\sqrt{\gamma}\alpha_i+ \dfrac{(\lambda k_c)^i}{\lambda\cdot(i-1)!} \\
\beta_i^-&=&\sqrt{\gamma}\alpha_i- \dfrac{(\lambda k_c)^i}{\lambda\cdot(i-1)!}
\end{eqnarray} 
and taking $\gamma$ sufficiently small so as to satisfy the following inequality:
\begin{eqnarray}
\sqrt{\gamma}\le \min_{i=1}^{n_\psi}\left[\dfrac{(\lambda k_c)^i}{2\alpha_i(\lambda\cdot(i-1)!)}\right] \label{suffsmall} 
\end{eqnarray} 
the following inequalities hold for $\beta_i^+$ and $\beta_i^-$:
\begin{eqnarray}
\beta_i^+\ge \dfrac{(\lambda k_c)^i}{\lambda\cdot(i-1)!}\quad;\quad \beta_i^-\le -\dfrac{(\lambda k_c)^i}{2\lambda\cdot\alpha_i((i-1)!)}
\end{eqnarray} 
With these inequalities, inequality (\ref{POh6}) implies:
\begin{eqnarray}
\dfrac{d\bar f}{dt}\le -\Bigl[\sum_{i=1}^{n_\psi}\dfrac{(\lambda k_c)^i}{(\lambda\cdot(i-1)!)}d^i(x,\rho)+{\relsize{-2}O(1/\rho)}\Bigr]\times\nonumber \\ \Bigl[\sum_{i=1}^{n_\psi}\dfrac{(\lambda k_c)^i}{2(\lambda\cdot(i-1)!)}d^i(x,\rho)+{\relsize{-2}O(1/\rho)}\Bigr] \label{POh687}
\end{eqnarray} 
Let us now show that inequality  (\ref{POh687}) implies that $\lim_{t\rightarrow \infty}\psi(x(t))=0$. Indeed, if this was not the case, then by the very definition of the dynamic on $\rho$ [see (\ref{ode2})] it comes that $\rho$ goes to infinity. This together with (\ref{POh687}) and the lower boundedness of $\bar f$ [Assumption \ref{ass1}] implies that $\lim_{t\rightarrow \infty} d(x,\rho)$=0 ($x$ converges to the set $\mathcal S_\rho$). But this implies by (\ref{eqass}) of assumption \ref{ass5} that $\psi(x)$ converges to $0$ which  contradicts the assumption. Now since $\psi(x)$ converges to $0$, the inequality (\ref{iju8}) together with the lower boundedness of $\bar f$ implies also that $g(x,\rho)$ converges to $0$.\ \\ \ \\ 
By now it has  been shown that provided that $\gamma$ is sufficiently small to satisfy (\ref{suffsmall}), the trajectory of $(x,\rho)$ converges to the following set
\begin{eqnarray}
\Bigl\{(x,\rho)\quad \vert\quad g(x,\rho)=0 \ \mbox{\rm and} \ \psi(x)=0\Bigr\} \label{theset} 
\end{eqnarray} 
It remains to prove that if $(x,\rho)$ belongs to the set defined by (\ref{theset}), then $x$ satisfies the KKT necessary conditions of optimality. Remember that these conditions require the existence of a vector $\mu\in \mathbb{R}^{n_c}$ such that the following conditions hold:
\begin{eqnarray}
\begin{array}{ll}
f_x(x)+\sum_{i=1}^{n_c}\mu_i\dfrac{\partial c_i}{\partial x}(x)=0 \cr
c_i(x)\le 0 &(\forall i\in \{1,\dots,n_c\})\cr 
\mu_i\ge 0 &(\forall i\in \{1,\dots,n_c\})\cr
\mu_i\times c_i(x)=0&(\forall i\in \{1,\dots,n_c\})
\end{array} 
\end{eqnarray} 
But $g(x,\rho)=0$ can be explicitly written as follows:
\begin{eqnarray}
f_x(x)+\rho\times m\sum_{i=1}^{n_c}\left[\max\{0,c_i(x)\}\right]^{m-1}\times \dfrac{\partial c_i}{\partial x}(x)
\end{eqnarray} 
which obviously shows that by taking $\mu$ such that:
\begin{eqnarray}
\mu_i:= \left\{ 
\begin{array}{ll}
 0& \mbox{\rm if $c_i(x)<0$}\\
 \rho\times m\times [c_i(x)]^{m-1}& \mbox{if $c_i(x)\ge 0$}
\end{array}
\right. \label{tgt543210} 
\end{eqnarray} 
the first KKT condition is satisfied by construction. The second condition ($c_i(x)\le 0$) results from $\psi(x)=0$. The third and the fifth conditions ($\mu_i\ge 0$ and $\mu_i\cdot c_i(x)=0$) result from (\ref{tgt543210}). This ends the proof. $\hfill \Box$ 
\ \\ \ \\ 
Note that if the summation in (\ref{ode1}) is performed with an infinite number of terms, the following corollary can be obtained:\\
\begin{corollary} \label{corollary1} 
Assume that some $\lambda>0$ is chosen. Consider the following system of differential equation:
\begin{eqnarray}
\dot x&=&-\exp\left[{\lambda\cdot g(x,\rho)}\right]\times \bar f_x(x,\rho) \label{odee1}\\
\dot\rho&=&\gamma\times \psi(x) \label{odee2}  
\end{eqnarray} 
{\bf If} Assumptions \ref{ass1}-\ref{ass5} hold then for sufficiently small $\gamma>0$, any asymptotic solution of (\ref{odee1})-(\ref{odee2}) satisfies the necessary conditions of optimality for the constrained optimization problem (\ref{defdeP}).  $\hfill \heartsuit$    
\end{corollary}
\ \\
Note that the result of Proposition \ref{prop1} holds for any initial condition that can be used to initialize the trajectory of (\ref{ode1})-(\ref{ode2}). The price to obtain such a global result lies in the use of the $q$-term summation  that premultiplies the gradient term $-\bar f_x(x,\rho)$ in (\ref{ode1}). The next proposition gives a weaker result that can nevertheless be preferable in some circumstances. In this weaker result, the convenient sufficiently small $\gamma$ would depend on the initial values of $x$ and $\rho$. \ \\ 
\begin{proposition}{\bf [A Simpler Weaker Result]} \label{prop2} 
Assume that some $\lambda>0$ is chosen. Consider the following system of differential equations:
\begin{eqnarray}
\dot x&=&-\bar f_x(x,\rho) \label{ode1bis}\\
\dot \rho&=&\gamma \times \psi(x) \label{ode2bis} 
\end{eqnarray} 
{\bf If} the following conditions hold:
\begin{enumerate}
\item Assumptions \ref{ass1}-\ref{ass5} are satisfied,
\item $f(\cdot)$ is proper (that is $\lim_{\|x\|\rightarrow \infty}f(x)=\infty$)
\end{enumerate}  
then for any initialization $(x_0,\rho_0)$, there is sufficiently small $\gamma>0$ such that the resulting asymptotic solution of (\ref{ode1bis})-(\ref{ode2bis}) satisfies the KKT necessary conditions of optimality for the optimization problem (\ref{defdeP}).  $\hfill \heartsuit$
\end{proposition}
\ \\
{\sc Proof}. Note that the result  can be obtained if one can show that everything behaves as if $n_\psi=1$ holds in (\ref{defdegrowthpsi}). Indeed, in this case $q=n_\psi=1$ can be used and (\ref{ode1}) is equivalent to (\ref{ode1bis}). This can be done using classical arguments that are typically used to derive semi-global results. More precisely, given the initial state $(x_0,\rho_0)$, define the following level set in $\mathbb{R}^{n}$:
\begin{eqnarray}
\mathcal V(x_0,\rho_0):=\Bigl\{x\ \vert \ f(x)\le 2\bar f(x_0,\rho_0)\Bigr\}
\end{eqnarray}
to which the initial value $x_0$ obviously belongs [because $f(x_0)\le \bar f(x_0,\rho)$ for all $\rho$]. Note that since $f$ is proper by assumption, the set $\mathcal V(x_0,\rho_0)$ is a compact set. Consequently, there is some sufficiently high $\bar\alpha_1$ such that the following inequality holds for all $(x_1,x_2)\in \mathcal V(x_0,\rho_0)$:
\begin{eqnarray}
\|\psi(x_2)-\psi(x_1)\|\le P(\|x_2-x_1\|)\le \bar\alpha_1\|x_2-x_1\| \label{ed6g5} 
\end{eqnarray} 
this means that as far as the trajectory remains in $\mathcal V(x_0,\rho_0)$, the result of Proposition \ref{prop1} can be used with $n_\psi=1$ therefore, there exists sufficiently small $\gamma$ such that the dynamics defined by (\ref{ode1}) with $q=1$ [which is the same as (\ref{ode1bis})] decreases the value of $\bar f(x,\rho)$. But this guarantees that the trajectory of $x$ remains in $\mathcal V(x_0,\rho_0)$. This implies that the inequality (\ref{ed6g5}) remains true and the result obviously follows. $\hfill \Box$     \ \\ \ \\ 
Note that such finite $\bar\alpha_1>0$ exists as long as the last inequality in (\ref{ed6g5}) is required only on the compact set $\mathcal V(x_0,\rho_0)$. The latter is defined in terms of the initial paire $(x_0,\rho_0)$. This is why the value of $\bar\alpha_1$ does depend on the initialization and may not exit globally. 
\subsection{General Comments}
\noindent Before getting to the examples section, it is worth mentioning that the results of the present section build a theoretical bridge between NLP and ODE algorithms in a rather systematic way and for a large class of problems. However, it must be underlined that although the following examples show rather efficient computational results, the integration of the resulting ODE may not be the more efficient way to solve the underlying optimization problems. This is because integration schemes try to reproduce high precision solution over the whole trajectories while from the NLP solution point of view, only the asymptotic trajectory matters. \ \\ \ \\ 
To this respect, the results of the present section can be used to derive gradient-based algorithms (fast gradient for instance \cite{Nesterov1983,Nesterov:04}) using the r.h.s of the ODE as extended gradient in the extended space of $(x,\rho)$ with $\bar f$ as cost function. By doing so, even certification results similar to the one proposed in \cite{Richter:2012} can be extended from the case where only saturations on the control input is used to the more general case of affine constraints on the state. This being said, no such efficiency-oriented development is done here focusing on the main theoretical contribution of the paper. \ \\ \ \\ 
On the other hand, another consequence of the theoretical result of the present section is the possibility to built electronic circuits that realize analogic ultra-fast integration of the ODE for a class of NLPs. This is briefly discussed in section \ref{circuits}. 
\section{Illustrative Examples} \label{secexamples} 
\subsection{Example 1: QP problems}
\begin{figure}
\begin{center}
\includegraphics[width=\textwidth]{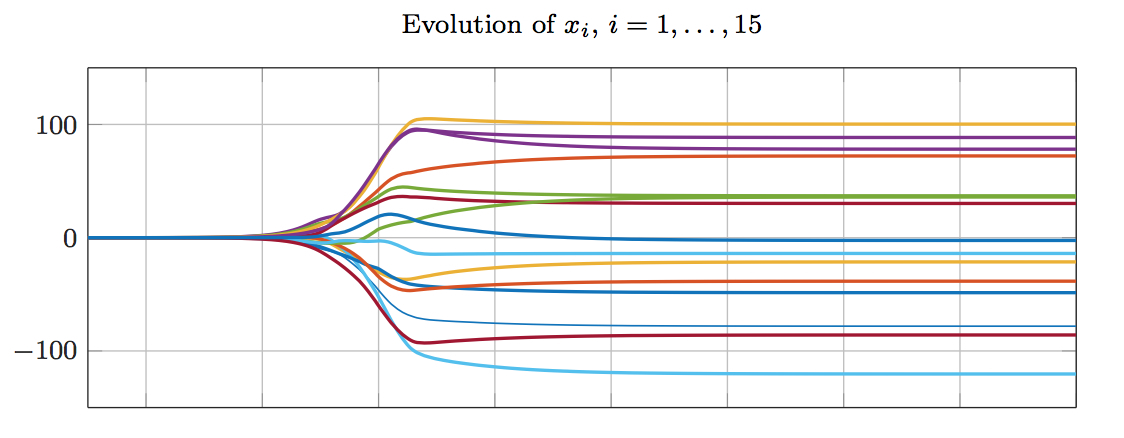}
\end{center} 
\caption{{\bf Example 1}. Typical behavior of of the components of $x$ on the trajectories of the ODE. Initial value $(0,0)\in \mathbb{R}^{n}\times \mathbb{R}$ is used.}\label{evolution_xi_qp}
\end{figure}

\noindent As a first examples let us consider the use of the ODE framework described in Proposition \ref{prop1} to solve Quadratic Programming (QP) problem with inequality constraints. This leads to the following instantiation of the cost function $f(x)$ and the constraints $c_i(x)$: 
\begin{eqnarray}
f(x)&=&\dfrac{1}{2}x^THx+F^Tx \label{defdeqp1} \\
c_i(x)&=&A_ix-B_i \quad ;\quad i=1,\dots,n_c \label{defdeqp2} 
\end{eqnarray}  
\begin{figure}
\begin{center}
\includegraphics[width=0.5\textwidth]{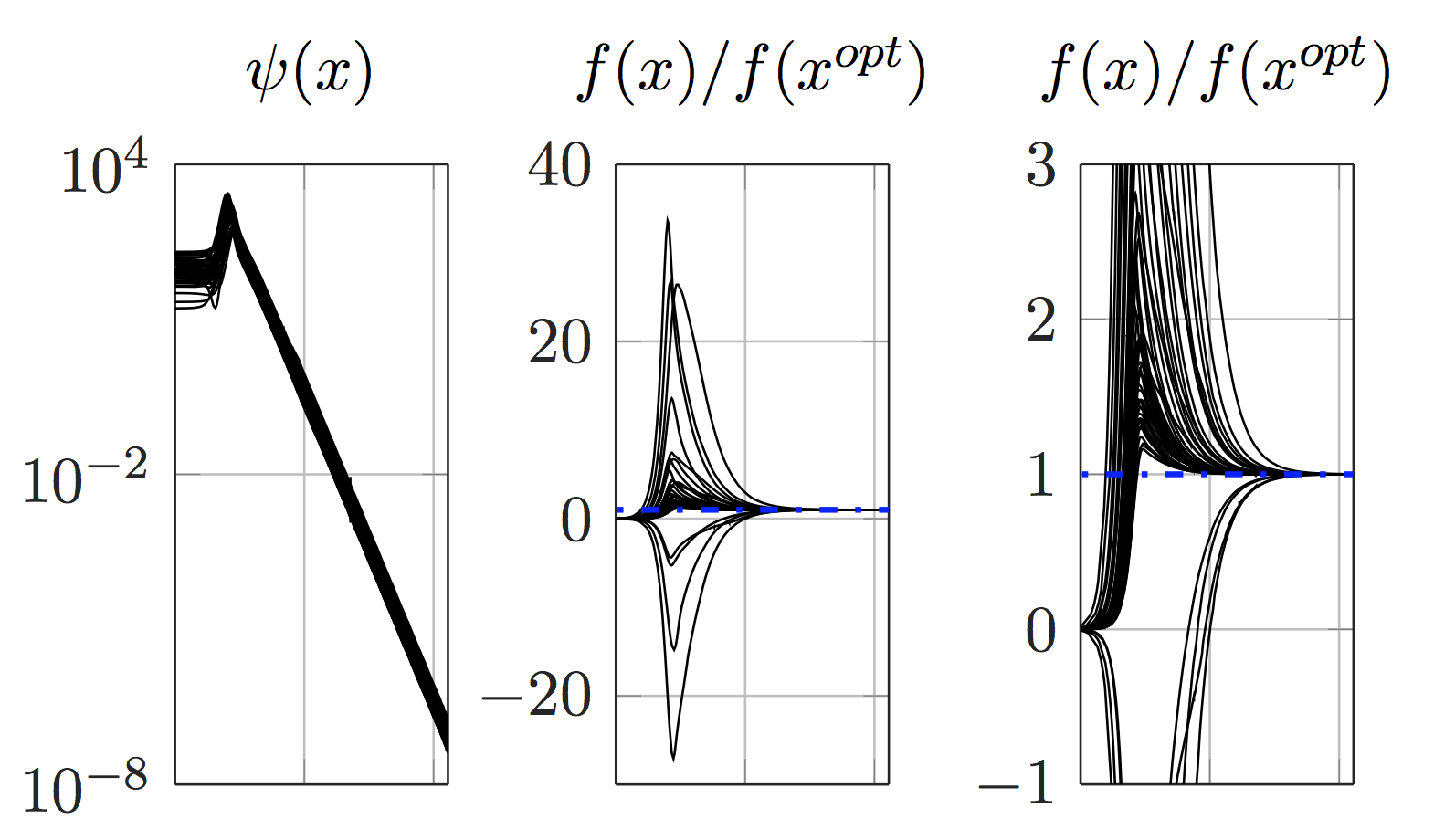}
\end{center}
\caption{{\bf Example 1}. Trajectories of $\psi(x)$ and $f(x)$ for $50$ different randomly generated problems. Note that the constraints are asymptotically satisfied ($\psi(x)\rightarrow 0$) and that the resulting costs converge towards the exact optimal value $f(x^{opt})$ (since $f/f^{opt}\rightarrow 1$). For all trials, the initial conditions $(0,0)$ is used. Initial values of $\psi$ shows initial strong violation of the constraints.} \label{validation_qp} 
\end{figure}
where $H\in \mathbb{R}^{n\times n}$, $F\in \mathbb{R}^{n\times 1}$, $A_i\in \mathbb{R}^{1\times n}$ and $B_i\in \mathbb{R}$. Now using $m=2$ to define the constraints-related weighting term:
\begin{eqnarray}
\psi(x)=\sum_{i=1}^{n_c}\bigl[\max\{0,A_ix-B_i\}\bigr]^2
\end{eqnarray} 
gives $n_\psi=2$ [see (\ref{defdegrowthpsi}) and (\ref{defdePol})]. Consequently, following Proposition \ref{prop1}, the following ODE is defined (taking $q=2$):
\begin{eqnarray}
\dot x&=&-\Bigl[1+\lambda\cdot \|\bar f_x(x,\rho)\|\Bigr]\times \bar f_x(x,\rho) \label{odeqp1} \\
\dot \rho&=&\gamma \times \sum_{i=1}^{n_c}\bigl[\max\{0,A_ix-B_i\}\bigr]^2 \label{odeqp2} 
\end{eqnarray}  
where 
\begin{eqnarray*}
\bar f_x(x,\rho)=
Hx+F+\rho\sum_{i=1}^{n_c}\max\{0,A_ix-B_i\}]\times A_i^T
\end{eqnarray*} 
This ODE is then integrated using the Matlab ODE15s stiff solver to get the solution of the original QP problem defined by (\ref{defdeqp1})-(\ref{defdeqp2}). Fifty randomly generated sets of matrices $\{H,F,A,B\}$ are generated leading to $50$ feasible QPs with $n=15$ unknown and $n_c=20$ constraints. The resulting ODEs (\ref{odeqp1})-(\ref{odeqp2}) are defined with the parameters $\lambda=10^{-4}$ and $\gamma=10^{-6}$. Figure \ref{validation_qp} shows the resulting trajectories of $\psi$ and the cost function normalized by the optimal cost value (computed using the standard Matlab QuadProg solver). All the trajectories are started from $x_0=0$ and $\rho=0$. The Figure clearly shows that the trajectories converge to the solutions of the problems as the constraints are satisfied and the cost function values converge toward the optimal values for all the generated problems. 
Figure \ref{evolution_xi_qp} shows a typical behavior of the system's trajectory starting from $(0,0)$ and converging towards the optimal values. The computation times shows a mean of $49\ ms$ with a variance of $3\ ms$ (Using Matlab on Mac PowerBook OSX, 2.8 GHz Intel Core i7 processor).

\subsection{Example 2: Analogic MPC Solvers} \label{circuits} 
\noindent Recall that Model Predictive Control (MPC) for linear time invariant systems of the form:
\begin{eqnarray}
\dot \xi=A\xi+Bu
\end{eqnarray} 
is based on the repetitive solution of a quadratic programming problem of the form:
\begin{eqnarray}
\min_{x\in \mathbb{R}^{n}} \Bigl[\dfrac{1}{2}x^THx+\bigl[f_0+F_1\xi\Bigr]^Tx\Bigr] \label{qpana1} 
\end{eqnarray} 
under the constraint:
\begin{eqnarray}
A_ix-\left[b_i^0+B_i\xi\right]\le 0\quad i\in \{1,\dots,n_c\} \label{qpana2} 
\end{eqnarray} 
where $x\in \mathbb{R}^{n}$ is the parameter vector that defines the control trajectory over the prediction horizon, namely:
\begin{eqnarray}
\begin{pmatrix}
u(k)\cr \vdots\cr u(k+N-1)
\end{pmatrix}=\Pi\cdot x
\end{eqnarray}  
for some appropriately chosen parametrization matrix $\Pi\in \mathbb{R}^{(Nn_u)\times n}$ where $n_u$ is the dimension of the control input $u$. Note that the only difference between (\ref{defdeqp1})-(\ref{defdeqp2}) and (\ref{qpana1})-(\ref{qpana2}) is that the affine term in (\ref{qpana1}) and the r.h.s of the inequalities (\ref{qpana2}) depends on the state of the controlled system $\xi$. For more details on MPC design, the reader can refer to \cite{Mayne:2000} \ \\ \ \\ 
Now applying Corollary \ref{corollary1} to the QP defined by  (\ref{qpana1})-(\ref{qpana2}) with a sufficiently small $\gamma$ for all initial conditions of interest, it comes that the QP solution (for a given $\xi$) can be obtained by integrating the following set of ODEs:
\begin{eqnarray}
\dot x&=&-\Bigl[Hx+f_0+F\xi\Bigr]+\nonumber \\ &+&2\rho\sum_{i=1}^{n_c}\Bigl[\max\bigl\{0,A_ix-b_i^0-B_i\xi\bigr\}\Bigr]\cdot A_i^T \label{tgfr4} \\
\dot \rho&=&\gamma \sum_{i=1}^{n_c}\Bigl[\max\bigl\{0,A_ix-b_i^0-B_i\xi\bigr\}\Bigr]^2 \label{tgfr5} 
\end{eqnarray} 
The idea is then to perform the integration through analogic circuits. The literature is very rich regarding the way transfer functions and more generally nonlinear differential relationships can be realized by analogic circuits (see \cite{Papazoglou:1997,Gunes:1997} and the references therein). Let us concentrate on the operations involved in (\ref{tgfr4})-(\ref{tgfr5}) to check that analogic realizations can be derived considering that $x$ is represented by a vector of currents (voltage options is also possible \cite{Yuanmao:2012}  although it is not discussed here):
\begin{itemize}
\item {\em Constant current sources} Note first of all that the constant terms $f_0$ and $b_i^0$  corresponds to tunable source of currents.
\item {\em State dependent current sources} The state dependent terms $F\xi$ and $B_i\xi$ are computed numerically and the results is assigned to another vector of current sources that remain constant during the integration step. This is the only numerical operation which determines the sampling rate of the resulting MPC controller.
\item {\em Linear combination of currents}. This concerns the terms $Hx$ and $A_ix$ and can be realized for instance using unity gain cells as shown for instance in \cite{Gunes:1997}.
\item {\em Current summation and substraction}. This concerns the realization of the sums $Hx+f_0+F\xi$ and $A_ix-b_i^0-B_i\xi$  and can be viewed as a particular instantiation of the previous item and can therefore be realized using unity gain cells. 
\item {\em Squaring signals}. This is necessary to compute the summated terms in (\ref{tgfr5}) and can be achieved for instance using the circuits proposed in \cite{Filanovski:1992,Hidayat:2008} or any later work containing more recent devices and architectures.
\item {\em Multiplication by $\rho$}. This operation can be realized through tunable gain or by using standard signal multipliers as the on proposed in \cite{Hidayat:2008}.   
\end{itemize} 
\ \\
Note that the time needed to analogically integrate (\ref{tgfr4})-(\ref{tgfr5}) is the time necessary to fill the corresponding circuit's capacitors. This time can be made extremely short (nano or even pico-seconds) if the problem is appropriately normalized so as to have its normalized solution components $\bar x_i$ scaled down so that they correspond to tiny capacitor voltages. \ \\ \ \\ 
Note that in the above presentation, the linear character of the controlled system plays no determinant role. Indeed, thanks to the possibility of signal multiplication, squaring and even the possibility to implement the square rooting of signals \cite{Filanovski:1992}, a wide class of ODE's that would be associated to the solution of a wide class of non quadratic constrained NLP can be analogically solved in extremely fast way. Moreover, the potential use of massively integrated circuit makes it possible to solve large scale problems in this way.\ \\ \ \\ 
Note finally that many of the above mentioned circuits can be realized using on-line assignable gains which makes it potentially possible to use the same circuits for many different problems. It remains however necessary to analyse the cost of such circuit design and realization which is beyond the scope of the present paper that studied the conceptual opportunities that are made possible by the ODE-related formulation of constrained optimization problems. 
\subsection{Example 3: Solving Nonlinear Mixed-Integer Optimization Problems}
\noindent In this section, presentation is done for the special case where all the decision variables are binary. The case where some decision variable can be continuous can be obtained easily with extra notational complexity. Consider the optimization problem given by:
\begin{eqnarray}
\min_{x\in \mathbb{R}^n} f(x)\quad \mbox{\rm under $\bar c_i(x)\le 0$ and $x_i\in \{0,1\}$} 
\end{eqnarray} 
for all $i\in \{1,\dots,\bar n_c\}$. \ \\ \ \\ 
It is well known that this problem can be put in the standard form (\ref{defdeP}) by transforming the binary  constraints $x_i\in \{0,1\}$ into standard constraints of the form
\begin{eqnarray}
x_i-x_i^2\le 0\quad -x_i\le 0\quad \mbox{\rm and}\quad x_i-1\le 0 
\end{eqnarray}  
which yields a number of inequality constraints $n_c=\bar n_c+3n$. Moreover, the integer $n_\psi$ that characterizes the growth of $\psi$ [see (\ref{defdegrowthpsi}) and (\ref{defdePol}) is given by $n_\psi=2m$ where $m$ is the exponent used in the definition (\ref{defdepsi}) of $\psi$.\ \\ \ \\ 
Using $m=2$ leads to the following definition of $\psi(x)$:
\begin{eqnarray*}
\psi(x)&=&\sum_{i=1}^{n_c}\Bigl[\max\{0,c_i(x)\}\Bigr]^2+\sum_{i=1}^{n}\Bigl[\max\{0,x_i-x_i^2\}\Bigr]^2\\
&+&\sum_{i=1}^{n}\Bigl[\max\{0,-x_i\}\Bigr]^2+\sum_{i=1}^{n}\Bigl[\max\{0,x_i-1\}\Bigr]^2
\end{eqnarray*}  
Now applying the result of Proposition \ref{prop1} suggests the solution of the combinatoric optimization problem can be done by integrating the following ODE:
\begin{eqnarray}
\dot x&=&-\Bigl[\sum_{i=1}^{4}\dfrac{(\lambda\cdot g(x,\rho))^{i-1}}{(i-1)!}\Bigr]\times \bar f_x(x,\rho) \label{odej1}\\
\dot \rho &=& \gamma\times \psi(x) \label{odej2} 
\end{eqnarray}  
Now obviously the admissible set is not a convex set and the presence of local minima is very likely. The following algorithm can be used to visit such local minima successively. In this algorithm a successively modified cost function $f^{(s)}(\cdot)$ where $f^{(0)}\equiv f$ is initialized to the original cost and where $s$ denotes the number of already visited local minima. The local minimum $x^{(s)}$ is found by integrating the ODE defined by the weighted function $\bar f^{(s)}(x,\rho)$ and its corresponding norm of the gradient $g^{(s)}(x,\rho):=\bar f_x^{(s)}(x,\rho)$, namely:
\begin{eqnarray}
\dot x&=&-\Bigl[\sum_{i=1}^{4}\dfrac{(\lambda\cdot g^{(s)}(x,\rho))^{i-1}}{(i-1)!}\Bigr]\times \bar f_x^{(s)}(x,\rho) \label{odehghg1}\\
\dot \rho &=& \gamma\times \psi(x) \label{odehghg2} 
\end{eqnarray}  
This is done starting from the initial condition $(x^{(s-1)},0)$ and integrating the ODE until some stopping conditions on both $\psi$ and $g$ are satisfied. Then a term is added to the cost function which makes the current solution $x^{(s)}$ inappropriate. This can be done by first defining a neighbor vector $z^{(s)}$ to $x^{(s)}$ such that:
\begin{eqnarray}
\|z^{(s)}-x^{(s)}\|_\infty=1\quad \mbox{\rm and}\quad   c(z^{(i)})\le 0 \label{yhg672} 
\end{eqnarray} 
The new cost function $f^{(s+1)}$ is now defined by:
\begin{eqnarray*}
f^{(s+1)}(x):=f^{(s)}(x)+(1+2f^{(s)}(z^{(s)}))\cdot \exp(\dfrac{\mu}{4}\|x-x^{(s)}\|^2)
\end{eqnarray*} 
Now for sufficiently high $\mu$, this new cost function is such that $x^{(s)}$ is no more a local minimum since 
$$\bar f^{(s+1)}(x^{(s)})=f^{(s)}(x^{(s)})+f^{(s)}(z^{(s)})+1>f^{(s)}(z^{(s)})$$
therefore, incrementing $s$ and firing the integration of the new resulting ODE (\ref{odehghg1})-(\ref{odehghg2}) starting from the initial condition $(x^{(s)},0)$ leads to a necessarily different minimum and so on. \ \\ \ \\ 
The only assumption that is implicitly assumed is that there always exists a neighbor vector $z^{(i)}$ that is admissible in the sense of (\ref{yhg672}). If this is not satisfied, less close $z^{(s)}$ can be searched provided that a deterministic generation process is defined.
\section{Conclusion and Future Work} \label{secconclusion} 
\noindent In this paper, it is shown that the solution of optimization problems with inequality constraints can be obtained by solving appropriately defined ODEs. In these ODEs, simultaneous dynamics are given to the decision variable as well as to the weight associated to the exact penalty term on the constraint violation. One of the major impacts of this result lies in the possibility to design analogic circuits that can quickly and physically integrate the corresponding ODEs. Pushing this latter idea towards a concrete realization is the obvious follow up of the present work. Another direction is to use the result to derive a fast gradient algorithm together with its associated certification bounds regarding the number of iterations that would be necessary to achieve a prescribed level of precision following the steps of \cite{Richter:2012} while including affine constraints that are not considered in \cite{Richter:2012}. This was not possible precisely because when standard fast gradient is used, only projection on the box-like set can be done while guaranteeing the decrease of the cost function. The formulation proposed in the present paper provide generalization of this property to an extended monotonically decreasing cost function provided that the r.h.s of the ODE is used as an extended gradient.  

\bibliographystyle{plain}
\bibliography{mybibfile}

\begin{thebibliography}{10}

\bibitem{Birgin:2012}
E.~Birgin and J.~M. Martinez.
\newblock Augmented lagrangian method with non monotone penalty parameters for
  constrained optimization.
\newblock {\em Computational Optimization and Applications}, 51(3):941--965,
  2012.

\bibitem{Byrd:99}
R.~Byrd, M.~Hribar, and J.~Nocedal.
\newblock An interior point algorithm for large-scale nonlinear programming.
\newblock {\em SIAM Journal on Optimization}, 9(4):877--900, 1999.

\bibitem{Yuanmao:2012}
Level-Shifting Multiple-Input Switched-Capacitor~Voltage Copier.
\newblock Y. yuanmao and k. w. e. cheng.
\newblock {\em IEEE Transaction on Power Electronics}, 27(2):828--837, 2012.

\bibitem{Filanovski:1992}
I.~M. Filanovski and H.~P. Baltes.
\newblock Simple cmos analog square-rooting and squaring circuits.
\newblock {\em IEEE Transactions on Circuits and Systems - I Fundamental Theory
  and Applications}, 39(4):312--315, 1992.

\bibitem{Gunes:1997}
E.~O. Gunes and F.~Anday.
\newblock Realization of voltage and current-mode transfer functions using
  unity gain cells.
\newblock {\em International Journal of Electronics}, 83(2):209--213, 1997.

\bibitem{Hidayat:2008}
R.~Hidayat, K.~Dejhan, P.~Moungnoul, and Y.~Miyanaga.
\newblock Ota-based high frequency {CMOS} multiplier and squaring circuit.
\newblock In {\em Proceedings of the 2008 International Symposium on
  Intelligent Signal Processing and Communication Systems}, 2008.

\bibitem{Leis:1988}
Jorge~R. Leis and Mark~A. Kramer.
\newblock Algorithm 658: Odessa\&\#8211;an ordinary differential equation
  solver with explicit simultaneous sensitivity analysis.
\newblock {\em ACM Trans. Math. Softw.}, 14(1):61--67, March 1988.

\bibitem{Mayne:2000}
D.~Q. Mayne, J.~B. Rawlings, C.~V. Rao, and P.~O. Scokaert.
\newblock Constrained model predictive control: Stability and optimality.
\newblock 36:789--814, 2000.

\bibitem{Nesterov1983}
Y.~Nesterov.
\newblock A method of solving a convex programming problem with convergence
  rate o (1/k2).
\newblock {\em Soviet Mathematics Doklady}, 27(2):372--376, 1983.

\bibitem{Nesterov:04}
Y.~Nesterov.
\newblock {\em Introductory Lectures on Convex Optimization}.
\newblock Springer-verlag, Berlin, Germany, 2004.

\bibitem{Papazoglou:1997}
C.~A. Papazoglou and C.~A. Karybakas.
\newblock Noninteracting electronically tunable {CCII}-based current-mode
  biquadratic filters.
\newblock {\em IEE Proc. Circuits Devices Syst.}, 144(3):178--184, 1997.

\bibitem{Richter:2012}
S.~Richter, C.N. Jones, and M.~Morari.
\newblock Computational complexity certification for real-time mpc with input
  constraints based on the fast gradient method.
\newblock {\em Automatic Control, IEEE Transactions on}, 57(6):1391--1403, June
  2012.

\bibitem{Voss199765}
D.~Voss and S.~Abbas.
\newblock Block predictor-corrector schemes for the parallel solution of
  {ODEs}.
\newblock {\em Computers \& Mathematics with Applications}, 33(6):65 -- 72,
  1997.

\end{thebibliography}
\end{document}